\documentclass[nohyperref]{article}

\usepackage{microtype}
\usepackage{graphicx}
\usepackage{subfigure}
\usepackage{booktabs} 

\usepackage{hyperref}

\usepackage[accepted]{icml2022_camera_ready}

\usepackage{amsmath}
\usepackage{amssymb}
\usepackage{mathtools}
\usepackage{amsthm}

\usepackage[capitalize,noabbrev]{cleveref}

\theoremstyle{plain}

\theoremstyle{definition}

\theoremstyle{remark}

\usepackage[textsize=tiny]{todonotes}

\icmltitlerunning{CheXplaining in Style: Counterfactual Explanations for Chest X-rays using StyleGAN}

\begin{document}

\twocolumn[
\icmltitle{CheXplaining in Style: Counterfactual Explanations for Chest X-rays using StyleGAN}



\icmlsetsymbol{equal}{*}

\begin{icmlauthorlist}
\icmlauthor{Matan Atad}{equal,tum}
\icmlauthor{Vitalii Dmytrenko}{equal,tum}
\icmlauthor{Yitong Li}{equal,tum}
\icmlauthor{Xinyue Zhang}{equal,tum}
\icmlauthor{Matthias Keicher}{tum}
\icmlauthor{Jan S. Kirschke}{tum,klinik}
\icmlauthor{Bene Wiestler}{tum,klinik}
\icmlauthor{Ashkan Khakzar}{tum}
\icmlauthor{Nassir Navab}{tum}
\end{icmlauthorlist}

\icmlaffiliation{tum}{Technical University of Munich, Germany}

\icmlaffiliation{klinik}{Klinikum rechts der Isar, Munich, Germany}

\icmlcorrespondingauthor{Matan Atad}{matan.atad@tum.de}

\icmlcorrespondingauthor{Yitong Li}{yi\textunderscore tong.li@tum.de}

\icmlkeywords{Machine Learning, ICML}

\vskip 0.3in
]



\printAffiliationsAndNotice{\icmlEqualContribution} 

\begin{abstract}
Deep learning models used in medical image analysis are prone to raising reliability concerns due to their black-box nature. To shed light on these black-box models, previous works predominantly focus on identifying the contribution of input features to the diagnosis, i.e., feature attribution. In this work, we explore \emph{counterfactual explanations} to identify what patterns the models rely on for diagnosis. Specifically, we investigate the effect of changing features within chest X-rays on the classifier's output to understand its decision mechanism. We leverage a StyleGAN-based approach (StyleEx) to create counterfactual explanations for chest X-rays by manipulating specific latent directions in their latent space. In addition, we propose EigenFind to significantly reduce the computation time of generated explanations. We clinically evaluate the relevancy of our counterfactual explanations with the help of radiologists. Our
code is publicly available.\footnote{\href{https://github.com/CAMP-eXplain-AI/Style-CheXplain}{https://github.com/CAMP-eXplain-AI/Style-CheXplain}}
\end{abstract}

\section{Introduction}
\label{sec:intro}

Chest X-ray, benefiting from its simple accessibility and fast availability, is currently one of the most common ways for the screening and diagnosis of a variety of thoracic diseases. Deep learning models have demonstrated promising potential for automated interpretation of chest X-rays at the level of practicing radiologists \cite{rajpurkar2017chexnet,chexpert,NIH}. However, the black-box nature of deep learning models raises concerns about their reliability in clinical applications \cite{khakzar2021explaining,khakzar2021towards}. It is essential to know which patterns the models rely on for diagnosis in the clinical routine.

\begin{figure}[t]
  \centering
   \includegraphics[width=1\linewidth]{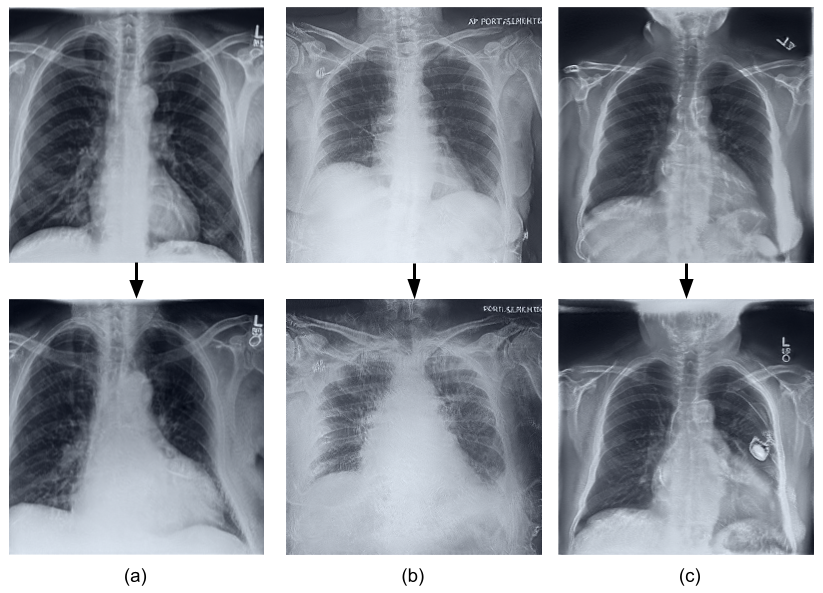}
   \caption{Some examples of the comparison of original chest X-ray images (first row) and their counterfactual explanations (second row) generated by our method. For each set of images, the emerging features in the counterfactuals are pathologically relevant: in (a) the width of the heart silhouette has been increased, corresponding to one of the main features of cardiomegaly; in (b) the appearance of the pleural recessus obstruction is a clear evidence of pleural effusion; in (c) a pacemaker was added, a demonstration of how the model learned to associate heart diseases (in this case cardiomegaly) with relevant indicators.}
   \label{fig:survey}
\end{figure}

\par

To interpret deep learning models in chest X-ray analysis, so far, most works leverage feature attribution (saliency) methods \cite{NIH,khakzar2021explaining,rajpurkar2017chexnet}. These  methods identify the contribution of input features to the diagnosis. Despite providing valuable information to the users, they only show which regions are important for the prediction on medical images, but there is ambiguity surrounding \emph{what} these features are. Some works \cite{wu2018deepminer,khakzar2021towards} provide further information regarding these features by analyzing neuron activation patterns on images with different concepts.

\par

However, there is an emerging avenue for neural network interpretation that is not explored in medical applications and is known as counterfactual explanations. Counterfactual explanations in medical image diagnosis models translate to identifying the feature changes required in the medical images in order to lead the model to a different diagnosis.

\par

To create counterfactual explanations, we need a method to extract visual features in images and change them in a semantic way. GAN-based models turn out to be an appropriate choice. Specific GAN structures can capture latent representations from the input data and control their features along these latent directions. \citet{lang2021explaining} presented a novel framework StyleEx to create a classifier-specific latent space and counterfactual explanations.

\par

In this paper, we explore counterfactual explanations for chest X-ray diagnosis models. We evaluate whether the counterfactual explanations are clinically relevant with the help of radiologists from our university hospital. We employed a method based on StyleEx \cite{lang2021explaining} and applied it to chest X-ray models to generate the counterfactual explanations. We improved the original method by factorizing the latent space instead of working on it directly, thus reducing the search time considerably.


\texttt{
\begin{figure*}[ht]
  \centering
   \includegraphics[width=1\linewidth]{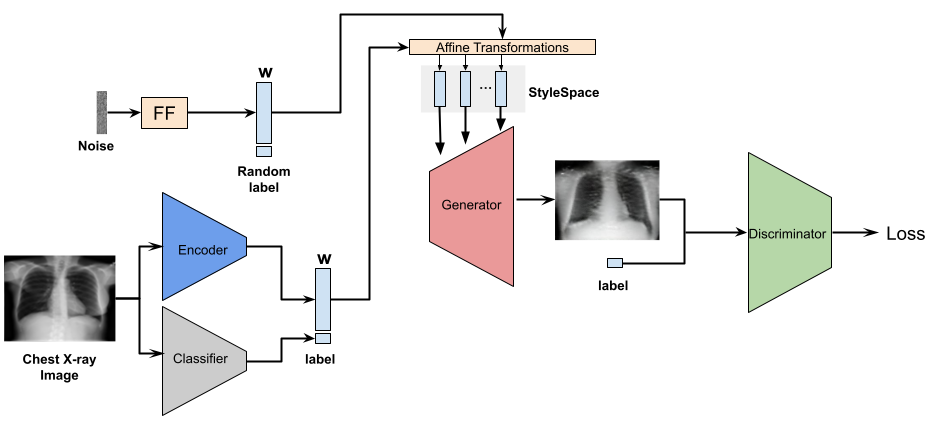}
   \caption{
   The pipeline captures classifier features in the StyleSpace, which are later used to generate counterfactual explanations. The whole architecture consists of a conditional StyleGAN and a pretrained frozen classifier, along with an encoder which allows mapping of images to the latent space. The Generator and Discriminator are conditioned on a label predicted by the classifier for the original input image. We use our EigenFind (\cref{alg:EigenFind}) to find the classifier specific directions in the learnt StyleSpace.}
   \label{fig:stylex}
\end{figure*}
}

\section{Method}
\label{sec:method}
In this section, we introduce the details of deploying the StylEx \cite{lang2021explaining} inspired methodology on chest X-ray models to generate counterfactual explanations. We further proceed with improving the style space search method to increase computational efficiency.

\par

Given an input X-ray image $x \in X$ and its matching classifier label $C(x)=y$, to create its counterfactual explanation, the aim is to change $x$ in a meaningful way, such that the changed image $\tilde{x}$ is as close as possible to $x$ but $C(\tilde{x})=\tilde{y}$ where $\tilde{y} \neq y$. 

\par
The method we use for counterfactual generation is based on StylEx proposed by \citet{lang2021explaining}.
Our implementation of StylEx is trained on the CheXpert dataset \cite{chexpert}. The architecture is comprised of a conditional StyleGAN2 \cite{karras2020analyzing}, a frozen pretrained Classifier and an Encoder (\cref{fig:stylex}). 

\par

We first pretrained a DenseNet \cite{iandola2014densenet} classifier on a Positive vs. Healthy binary setting per pathology in the dataset. Both the Generator and the Discriminator will then be conditioned on the class labels $y$ predicted from the classifier by concatenating an embedding of the image labels to their inputs. The encoder is based on the Discriminator architecture while removing the batch-normalization layer. The main purpose of the encoder is to allow for mapping of any kind of image into the latent space, by which we will be able to create counterfactuals for real images. 

\par

We trained the entire architecture at once, passing in each iteration both latents originating in noise and in encoded images. Training along with the raw noise input will help us to maintain the optimal training direction of the conditional StyleGAN. The trained StyleSpace will capture the features that are decisive for the classifier's prediction, the most significant features of which will be later detected and extracted using specific searching algorithm for counterfactual generation.  

\par

We propose our algorithm EigenFind (\cref{alg:EigenFind}) for a more efficient counterfactual search. In the previous paper, \citet{lang2021explaining} presented the AttFind algorithm which iterates over all coordinates in the StyleSpace while changing them one by one, searching for coordinates with the largest affect on the classifier decision. We factorize the StyleSpace with PCA \cite{shen2021closed} and modify the algorithm to iterate over Eigenvectors instead.

\begin{algorithm}[t]
   \caption{EigenFind}
   \label{alg:EigenFind}
\begin{algorithmic}
   \STATE {\bfseries Input:} Classifier $C$, Encoder $E$, Generator $G$, number of Eigenvectors to consider $k$, Degree of change $d$, Images $X$ classified as $y$ \\
   \STATE {\bfseries Return:}  Counterfactuals $X_{explained}$ \\
   \STATE
   \STATE{$X_{explained} \gets \varnothing$} \\
   \STATE{$V_{max} \gets \varnothing$}
   \STATE{$V \gets PCA(G)[1:k]$}
   \FOR{$v$ in $V$}
       \FOR{$x$ in $X$}
            \STATE{$\tilde{x} \gets G(E(x) + d * v)$}
            \STATE{$\delta[x, v] \gets C(\tilde{x}) - C(x)$}
       \ENDFOR
       \STATE{$\bar{\Delta}[v] = \frac{1}{|X|}\sum_{x \in X}{\delta[x, v]}$}
   \ENDFOR
   \REPEAT
        \STATE{$v_{max} \gets \operatorname*{argmax}_{v} \bar{\Delta}$}
        \FOR{$x$ in $X$}
            \STATE{$\tilde{x} \gets G(E(x) + d * v_{max})$}
            \IF{$C(\tilde{x}) = \bar{y}$}
                \STATE{$X_{explained} = X_{explained} \cup \tilde{x}$}
                \STATE{$X = X \setminus x$}
            \ENDIF
        \ENDFOR
    \STATE{$V_{max} = V_{max} \cup v_{max}$}
    \STATE{delete $\bar{\Delta}[v_{max}]$}
   \UNTIL{$|X|=0$ \textbf{or} $|\bar{\Delta}|=0$}
\end{algorithmic}
\end{algorithm}

\par

In EigenFind, for images $X$ classified as $y$, we calculate how moving their latents in the direction of each of the top $k$ StyleSpace Eigenvectors affects the classifier decision\footnote{Both positive and negative directions are evaluated, but the negative directions are omitted here for brevity.}. Next, we follow \citet{lang2021explaining} and estimate the most significant Eigenvectors by calculating the average difference between the classifier logit before and after the change on all input images $X$. Finally, for each image, we find which of the most significant Eigenvectors is able to flip the image label to $\tilde{y}$. The resulting image $\tilde{x} \in X_{explained}$ is the counterfactual.

\par

The time complexity of EigenFind is $\mathcal{O}(kn)$, where $n$ is the number of images and $k$ is the number of top Eigenvectors we consider. The time complexity of AttFind \cite{lang2021explaining} is also linear $\mathcal{O}(mn)$, where $m$ is the number of channels in the StyleSpace feature map. In practice, $m$ is determined by the StyleGAN architecture based on the size of the input image (for $256 \times 256$ $m = 3040$\footnote{We do not consider the StyleGAN ToRGB layers as part of the StyleSpace given to the AttFind search, since these are shown to affect only the output image color \cite{wu2021stylespace}.}) and we chose $k = 8$. Since $k \ll m$ the running time is considerably reduced.


\section{Experiments \& Evaluation}
\label{sec:experiments}

We trained the pipeline separately for three common thoracic pathologies in the CheXpert dataset \cite{chexpert}: Cardiomegaly, Pleural Effusion and Atelectasis. Based on our EigenFind, we then generated counterfactual explanations for these pathologies. Some examples are shown in \cref{fig:survey} - the first row demonstrates the original healthy chest X-ray images, while the second row concludes the corresponding generated counterfactuals. From each pair of images, the features emerging in the counterfactuals are representative of the main features of each disease, which visually affirms the pathological-relevance of the features found by EigenFind.

\par

Furthermore, in order to evaluate whether the counterfactual explanations found by our method are indeed clinically relevant, we cooperated with radiologists from our university hospital. They helped to diagnose which features changed in counterfactuals vs. the originals for these three pathologies (\cref{tab:table2}).

\par

In our evaluation setting, the radiologists first listed the main features and possible secondary findings during diagnosis of each disease. After randomly selecting 10 images that have been classified as \emph{Healthy} samples (i.e., originals), we separately moved their latent representations in the direction of each of the three most significant Eigenvectors, which were obtained with EigenFind (\cref{alg:EigenFind}). After each movement, the radiologists evaluated whether the previously listed disease features existed in the newly generated images (i.e., counterfactuals) or not. 

\par

The evaluation results are demonstrated in the last three columns of \cref{tab:table2}. The results indicate that most of the main features could be spotted in the counterfactuals generated by our EigenFind, thus indicating that our most significant Eigenvectors can help with identifying clinical-relevant features. By doing so, we can not only easily spot which regions are crucial for the classifiers to predict different pathologies, but more importantly, we can also understand \emph{what} these determined features are, by comparing the vivid changes between the original images and their counterfactuals.

\par

In addition, we compared the ability to explain images (i.e to create counterfactuals) of our EigenFind algorithm with \citet{lang2021explaining} AttFind. In \cref{tab:table1} our method achieves results on par with \citet{lang2021explaining}, while requiring a fraction of the search time on a Tesla P100 GPU.

\par

In our experiments, the StyleGAN2 pipeline was trained with an Adam optimiser for $40K$ iterations with a batch size of $32$ on images of size $256 \times 256$. The learning rate was set to $0.0016$ for the Generator, $0.0018$ for the Discriminator and $0.002$ for the Encoder. Path length regularization was applied every $4$ epochs for the Generator and $R1$ regularization was applied every $16$ epochs for the Discriminator. For the evaluation of the counterfactual search algorithms, we took $600$ random images for each pathology. For EigenFind we considered the top $k = 8$ Eigenvectors and a degree of change of $d = 10$.

\begin{table}[t]
  \caption{Comparison of the percentage of explained images for each of the three pathologies in CheXpert \cite{chexpert} (i.e., ones for which a counterfactual could be created), along with the search time between the two algorithms: AttFind by \citet{lang2021explaining} and our EigenFind. Both algorithms achieved comparable results on counterfactual generation, while our method reduced the searching time considerably.}
  \vskip 0.15in
  \centering
  \begin{tabular}{@{}lcc@{}}
    \toprule
     & AttFind & EigenFind \\
    Pathology & \citet{lang2021explaining} &  Ours \\
    \midrule
    Atelectasis & $94\%$ & $94\%$ \\
    Cardiomegaly & $96\%$ & $95\%$ \\
    Pleural Effusion & $94\%$ & $91\%$ \\
    \hline
    Search Time & 12 hours & 5 minutes \\
  \end{tabular}
  \label{tab:table1}
\end{table}


\begin{table}[th]
  \caption{Radiologists' evaluation of our generated counterfactuals. Common disease features and secondary findings for each of the three pathologies are listed by the radiologists in the first column. Then the radiologists evaluated if the corresponding features existed, after moving the latent representation of a set of 10 Healthy images in the direction of the 3 most significant Eigenvectors respectively (listed as the last three columns).}
  \vskip 0.15in
  \centering
  \begin{tabular}{l|c|c|c}
    \toprule
    Features of Cardiomegaly & $v_{17}$ & $v_{2}$ & $v_{18}$ \\
    \midrule
    Increased cardiothoracic ratio & \checkmark & \checkmark & \checkmark \\
    \textit{Secondary findings:} & & & \\
    \hspace{3mm} Reduced lung tissue opacity & \checkmark  &  & \\
    \hspace{3mm} Pleural Effusion & \checkmark & \checkmark &  \\
    \hspace{3mm} Pacemaker &  &  &  \checkmark \\
    \hspace{3mm} Older patients & & \checkmark & \\
    
    \toprule
    Features of Pleural Effusion & $v_{15}$ & $v_{7}$ & $v_{12}$ \\
    \midrule
    Obstruction of the pleural recessus & \checkmark  & \checkmark  & \checkmark \\
    Opaque lower lungs & & & \checkmark \\
    \textit{Secondary findings:} & & & \\
    \hspace{3mm} Increased cardiac diameter & \checkmark  &  & \\
    \hspace{3mm} Fluid overload & \checkmark  &  & \\
    \hspace{3mm} Pneumonia &  & \checkmark & \\
    
    \toprule
    Features of Atelectasis & $v_{15}$ & $v_{7}$ & $v_{12}$ \\
    \midrule
    Mediastinal shift & \checkmark & \checkmark & \checkmark \\
    Wide barrel-like thorax & \checkmark & &\checkmark \\
    \textit{Secondary findings:} & & & \\
    \hspace{3mm} Pleural Effusion & \checkmark & \checkmark  & \checkmark \\
    \hspace{3mm} Infiltration & \checkmark & \checkmark &   \\
    \hspace{3mm} Older patients & \checkmark &  & \\
    
    \hline
  \end{tabular}

  \label{tab:table2}
\end{table}
\section{Conclusion}
\label{sec:conclusion}

To address the concerns regarding the explainability of deep learning models in medical image analysis, we explored the chest X-ray domain and investigated counterfactual explanations to identify the feature changes in chest X-ray images that can lead the classifier to a different diagnosis. 

\par

To create such counterfactual explanations, we leveraged a StyleGAN-based approach by manipulating specific latent directions in the pre-trained latent space of the generator. The newly generated images after the latent space manipulation become our counterfactuals. We also propose the EigenFind algorithm to significantly reduce the computation time for counterfactuals generation by factorizing the latent space with PCA, and working on the Eigenvectors instead.

\par

Furthermore, we evaluated whether such classifier-decisive features spotted by our EigenFind algorithm are clinically relevant with the help of radiologists. The results demonstrate that the most significant Eigenvectors obtained from EigenFind are able to help with identifying clinical-relevant features in chest X-rays. The feature changes in the generated counterfactuals are in accordance with the main diagnosing features for common thoracic diseases. In addition, the model also learned to associate thoracic diseases with relevant indicators such as pacemaker and age.

\bibliography{egbib}
\bibliographystyle{icml2022}

\end{document}